# INFLUENCE OF CREAMING AND RIPENING ON THE AGGREGATION RATE OF NON-IONIC DODECANE-IN-WATER NANOEMULSIONS

Eliandreína Cruz-Barrios and German Urbina-Villalba*

Instituto Venezolano de Investigaciones Científicas (IVIC), Centro de Estudios Interdisciplinarios de la Física, Apartado 20632, Edo. Miranda 1204, Venezuela. Email: guv@ivic.gob.ve

**Abstract** The possible influence of creaming during the measurement of the aggregation rate of dodecane-in-water nanoemulsions stabilized with Brij 30 is explored. For this purpose additional emulsions made with a neutral-buoyancy oil (bromo-dodecane) and mixtures of dodecane and Br-dodecane with squalene were synthesized. It is concluded that when the effect of ripening is suppressed, the influence of buoyancy on the evaluation of the flocculation rate is negligible. In the absence of squalene, ripening is present, and a sizeable difference in the flocculation rate of dodecane and bromo-dodecane is observed. However, this difference is not caused by the effect of gravity.

## INTRODUCTION

Oil-in-water nanoemulsions are thermodynamically unstable dispersions of two immiscible liquids with a drop size between 20-500 nm. As typical emulsions, these systems are destabilized by several mechanisms like creaming, Ostwald ripening, flocculation, and coalescence. These processes are interrelated and occur simultaneously making difficult to identify the origin of the instability of the dispersions [Mendoza, 2013a]. However, due to the small size of the drops, one of the main mechanisms of instability is the molecular exchange of oil between droplets, known as Ostwald ripening. According to the theory of Lifshitz, Slezov and Wagner [Lifshitz, 1961; Wagner, 1961] the rate of ripening ($V_{OR}$) can be quantified in terms of the linear increase of the cubic average radius ($R_c$) of the emulsion as a function of time:

$$V_{OR} = \frac{dR_c^3}{dt} = \frac{4}{9}\alpha D_m C_\infty \quad (1)$$

Here $D_m$, $C_\infty$ and $\alpha$ stand for the diffusion constant of the oil molecules in water, their solubility, and the capillary length of the drops, defined as:

$$\alpha = \frac{2\gamma V_m}{\tilde{R}T} \quad (2)$$

Where: $\tilde{R}$ is the universal gas constant, $T$ the absolute temperature, $\gamma$ the interfacial tension of the drops, and $V_m$ the molar volume of the oil.

The fact that the rate of ripening is determined through the variation of the average radius of the emulsion is very unfortunate, since this radius changes with every process of destabilization of the system. In fact, the change of the average radius and the drop size distribution (DSD) as a function of time, is the best estimation of the overall stability of the system [Mendoza, 2013a]. In any event, it is clear from Eq. (1) that the rate of ripening depends markedly on the solubility of the oil. Moreover, it is more pronounced for small drops, since the capillary length is generally of the order of nanometers, and according to the Kelvin equation, the solubility of an oil drop of radius R, C(R), is inversely proportional to its radius [Kabalnov, 2001]:

$$C(R) = C_\infty \exp\left(\left(\frac{1}{R}\right)\frac{2\gamma V_m}{\tilde{R}T}\right) =$$
$$= C_\infty \exp\left(\frac{\alpha}{R}\right) \approx C_\infty\left(1 + \frac{\alpha}{R}\right) \quad (3)$$

Kabalnov [Kabalnov, 2001] further showed, that the rate of ripening can be substantially reduced adding a very insoluble substance to the matrix oil. For a system to be thermodynamically stable with respect to Oswald ripening, the initial molar fraction of the insoluble compound ($x_{02}$) has to be greater than:

$$x_{02} > \alpha/R_0 \quad (4)$$

Where $R_0 = R_c(t = 0)$ is the initial particle radius. Con-dition (4) cannot always been achieved, and in most cases, only metastable (kinetic) stability can be attained:

$$\alpha/R_0 > x_{02} > \alpha/3R_0 \quad (5)$$

In particular, we had proven that the addition of 10% of squalene ($C_{30}$) to dodecane, leads to the fulfillment of Eq. (5) [Cruz-Barrios, 2014].

As in macroemulsions, the kinetic stability of nanoemulsions with respect to flocculation and coalescence is achieved by the addition of a surface active agent. This produces a repulsive interaction between the drops which depends on the chemical structure of the surfactant. Ionic surfactants create electrostatic barriers against flocculation that can be screened with the addition of salt. Non ionic surfactants provide steric barriers between the drops which are inversely proportional to the temperature of the system [Capek, 2004; Lozsán, 2005, 2006].

The rate of flocculation is usually assessed in solid dispersions measuring the rate of doublet formation ($k_{11}$). This variable is proportional to the initial variation of the turbidity ($\tau_{exp}$) as a function of time [Lips, 1971; 1973]:

$$\left(\frac{d\tau_{exp}}{dt}\right)_0 = 230\left(\frac{dAbs}{dt}\right)_0 = \left(\frac{1}{2}\sigma_2 - \sigma_1\right)k_{11}n_0^2 \quad (6)$$

Where $\sigma_1$ and $\sigma_2$ are the optical cross sections of a single drop and a doublet, respectively, $t$ is the time, and $Abs$ stands for the absorbance of the dispersion.

More recently we implemented a procedure that allows the evaluation of a unique flocculation rate for the whole process of aggregation [Rahn-Chique, 2012]:

$$\tau_{teo} = n_1\sigma_1 + x_a\sum_{k=2}^{k_{max}}n_k\sigma_{k,a} + x_s\sum_{k=2}^{k_{max}}n_k\sigma_{k,s} \quad (7)$$

Here: $\sigma_{k,a}$ and $\sigma_{k,s}$ represent the optical cross sections of an aggregate of size $k$ and the one of a spherical drop of the same volume. $x_a$ stands for the fraction of collisions that only lead to the flocculation of the drops, $x_s$ is the fraction of collisions that results in their coalescence, and $n_k$ is the number density of aggregates of size $k$ existing in the dispersion at time t [Smoluchowski, 1917]:

$$n_k = \frac{n_0(k_{FC}n_0t)^{k-1}}{(1+k_{FC}n_0t)^{k+1}} \quad (8)$$

In Eq. (8) $n_0$ represents the total number of aggregates at time t = 0 ($n_0 = \sum n_k(t=0)$) and $k_{FC}$ is an average aggregation-coalescence rate. In the absence of mixed aggregates: $x_s = (1-x_a)$. The values of $k_{FC}$ and $x_a$ result from the fitting of Eq. (7) to the experimental variation of the turbidity as a function of time $\tau_{exp}(t)$.

In regard to, a simplified formula can be obtained from the balance of the gravity and the buoyancy force:

$$V_g = \frac{2R_i^2\Delta\rho g}{9\eta} \quad (9)$$

Where $V_g$ is the creaming rate, $\Delta\rho$ the density difference between the external and the internal phase, g is the gravity force (9.8 m/s$^2$), and $\eta$ is the viscosity of the water phase. More elaborate formulas for the creaming rate involve the volume fraction of internal phase $\phi$ (Migration software [Formulaction, 2002]) but give essentially the same result:

$$V_g = \frac{2R_i^2\Delta\rho g}{9\eta}\left[\left|1-\phi\right|/\left|1+4.6\phi/(1-\phi)^3\right|\right] \quad (10)$$

It is therefore noteworthy that while these equations predict that a column of emulsion of 10 cm will need 146 days in order to form cream, typical nanoemulsions –as the one synthesized in this work- show a maximum period of separation of 3 days. This evidences that either the drops are growing appreciably more rapidly than expected due to the influence of Ostwald ripening, or they are flocculating irreversibly.

The purpose of this work is to appraise the influence of creaming during the measurement of the flocculation rate of a non-ionic nanoemulsion. Unlike similar evaluations on ionic-nanoemulsions for which a 60-second period suffices, the ones of non-ioinic dispersions require about 15 hours (at T = 25°C) due to their greater stability. Hence, it might be expected that some aggregates cream out of the range of the detector during the measurements affecting the turbidity of the emulsion. However, this is uncertain, since the light source strikes the container at half its height. Hence, as long as there are enough particles in the lower half of the container during the measurement, aggregates that cream out of the scattering zone, are partially substituted by those (that coming from the lower half of the container) cream into the sample region.

In order to quantify the effect of buoyancy on the evaluation of the flocculation rate, we use here both dodecane ($C_{12}$) and bromo-dodecane ($Br-C_{12}$) nanoemulsions, as well as mixtures of these two components with squalene ($C_{30}$).

**EXPERIMENTAL DETAILS**

2.1 Materials

Dodecane (Aldrich, 99%, 0.75 g/ml), bromododecane (Aldrich, 99%, 1.04 g/ml), squalene (Aldrich, 99%, 0.81 g/ml), and Brij 30 ($C_{12}H_{25}(OCH_2)_n$, n~4, Aldrich, 0.95g/ml) were used as received. Water was distilled and deionized (1.1 $\mu S/cm^{-1}$ at 25ºC) with a Simplicity Millipore purifier.

2.2 Nanoemulsion synthesis and characterization

Nanoemulsions were obtained by a stepwise addition of water to an oil surfactant solution [Forgiarini, 2001], using the phase inversion temperature method [Shinoda, 1969; Wang, 2008]. The phase inversion temperature (PIT) was determined by conductivity measurements (Crison, model 525 with a platinum electrode). A pronounced fall in the conductivity is observed at the temperature in which the emulsion changes from oil-in-water to water-in-oil. The PIT was found equal to 32.8 ºC for $C_{12}$ and 36.5 ºC for $Br-C_{12}$. The samples were prepared using an electrolytic solution (NaCl $10^{-2}$ M) instead of pure water in order to emphasize the change in the conductivity of the emulsions at the phase inversion temperature.

Two sets of nanoemulsions were synthesized based on the composition of the oil: a) dodecane ($C_{12}$) and b) bromododecane ($Br-C_{12}$). These pure systems correspond theoretically to "finite" and "null" buoyancy experiments, respectively. Additionally, mixtures of $C_{12}$ and $Br-C_{12}$ with several percentages of squalene ($C_{30}$) between 2 and 15 % (weight percentages) were used to diminish the influence of Ostwald ripening on the destabilization [Higushi and Misra, 1962]. This allows separating its contribution from the effect of creaming. The PIT of the mixed nanoemulsions varied between 36-38 ºC, producing an initial average diameter of the drops between 127-131 nm with a polydispersity of 23 - 25%. The size distribution and polydispersity were determined using a particle size analyzer Beckman Coulter LS 230. The size distribution was unimodal in all the cases. A typical DSD is shown in Figure 1.

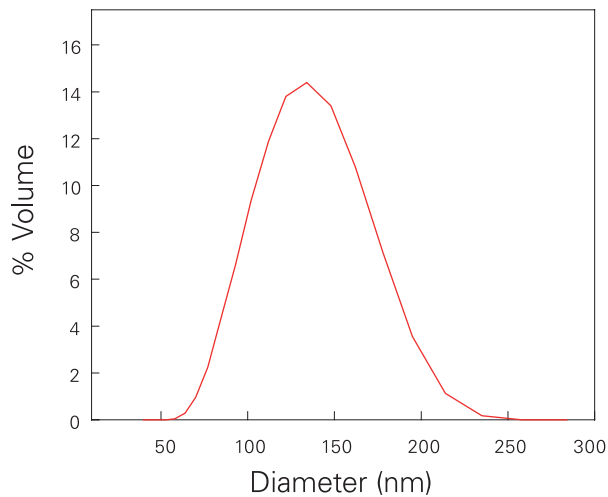

Figure 1: Typical initial drop size distribution of the dodecane-in-water nanoemulsions studied

2.3 Emulsions stability with respect to creaming

The QuickSCAN apparatus (Coulter) was used to evaluate the variation of the turbidity of the dispersion as a function of its height in the container. The evolution of the average radius was determined using a Brookhaven BI-200SM goniometer at a 90º fixed angle, at a 633-nm wavelength. Flocculation rates were determined from the variation of the absorbance as a function of time using an UV-visible spectrophotometer (Shimatzu 8800) at $\lambda$ = 500 nm. Different degrees of destabilization were induced increasing the temperature of the emulsions between 25-90 ºC. The final particle density of the nanoemulsions after dilution was $n_0$ = 1.3 x $10^{17}$ part/$m^3$ ($\phi$= $10^{-4}$). The surfactant concentration was adjusted to 6.05 x$10^{-5}$ M, just below the critical micelle concentration of Brij30 (CMC = 6.4 x $10^{-5}$ M) [Rosen, 1982] in order to avoid possible extraneous signals due to micellization [Cruz-Barrios, 2014].

**RESULTS AND DISCUSSION**

Figures 2 and 3 show the patterns of transmittance (%T) and backscattering (%BS) for the concentrated nanoemulsions of finite ($C_{12}$) and null buoyancy ($Br-C_{12}$) during a period of 24 hours. In the case of dodecane there is a progressive increase in the %T at the bottom of the sample with time, due to its creaming. This creaming progressively decreases as the amount of squalene in the sample increases, suggesting that the growth in the particle size is at least partially caused by Oswald ripening. Since $C_{30}$ is substantially lighter than dodecane, the creaming reappears when the amount of squalene increases beyond 10%. A similar phenomenon occurs in the case of Br-$C_{12}$. However, since the density of this substance is almost equal to the density of water, the variation of the transmittance is uniform throughout the sample, but still indicates a progressive change in the particle size. When a small amount of $C_{30}$ is added to the oil, the increase in the particle size apparently disappears, reappearing at 15% $C_{30}$. Since the emulsions used for these turbidity measurements are concentrated ($\phi$ = 0.25), the information obtained from the backscattering of light is substantially more informative. Notice that the amount of squalene needed to lessen the increase in the particle size is higher for Br-$C_{12}$ than for $C_{12}$ due to the smaller solubility of the latter. However, a comparison between the figures corresponding to 10% $C_{30}$ demonstrate, that the increase in the particle size is slower for Br-$C_{12}$ than for $C_{12}$ suggesting that once the effect of ripening is stopped, it is the difference in the density of the substances which drives the change in %BS. In fact, the effect of creaming at the top of the container is clear for $C_{12}$ and $C_{12}$ + 10% $C_{30}$, but it is only evidenced in the case of bromododecane at 15% $C_{30}$.

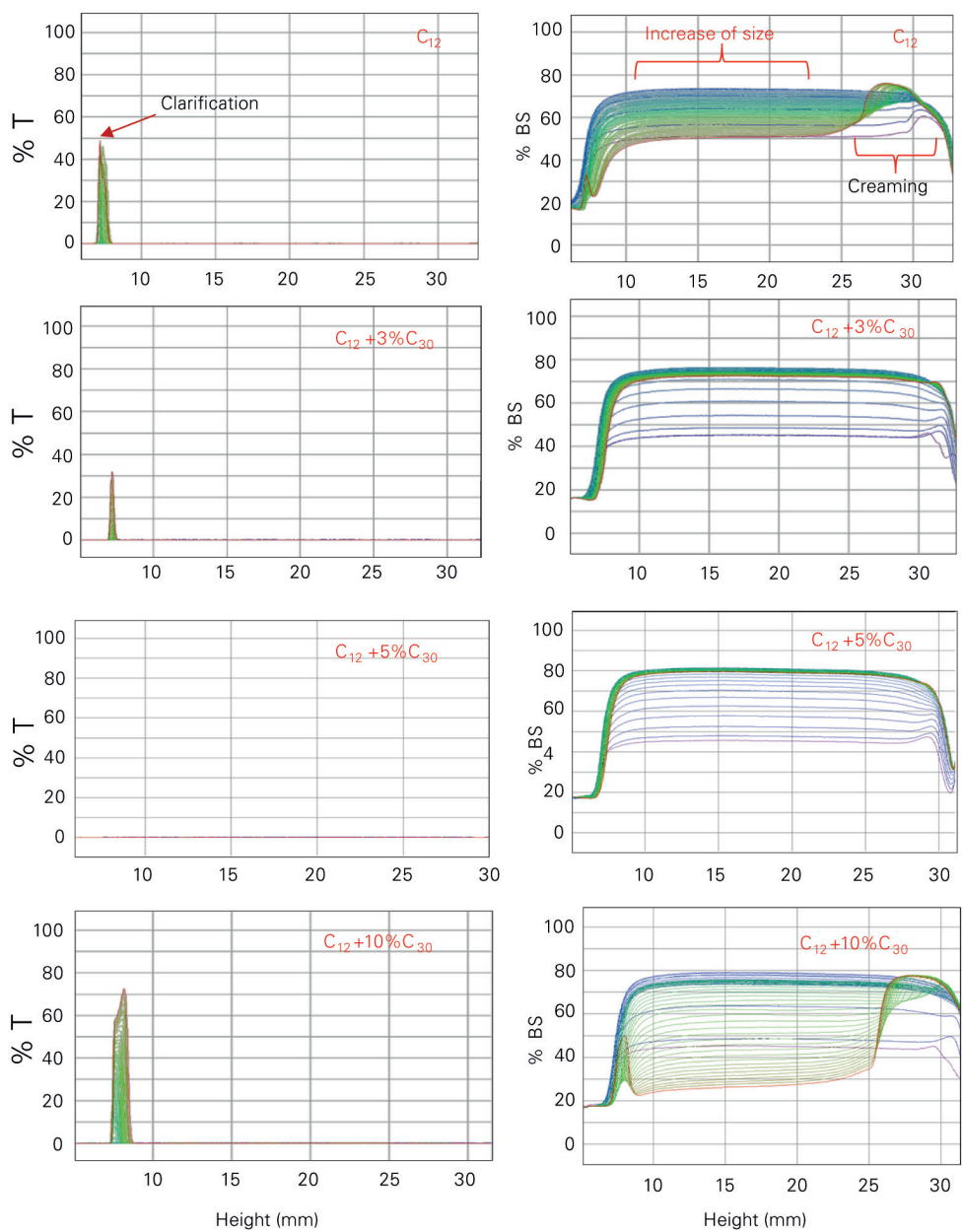

Figure 2: Temporal evolution of the percentage of Transmittance (%T) and Backscattering (%BS) of light scattered by drops of (dodecane ($C_{12}$) + squalene ($C_{30}$)) suspended in water as a function of their height in the sample vessel.

Figure 4 shows the variation of the cube average radius as a function of time for: $C_{12}$ and Br-$C_{12}$ obtained from the goniometer. The dense sampling of the system produces curves with a significant statistical noise. An "initial" and a "terminal" slope of the curves were calculated in order to have an estimation of the closeness of these values to the ones of the so-called stationary regime. As time progresses the slope of the curves diminishes but it only changes in orders of magnitude (from $10^{-27}$ to $10^{-29}$ m$^3$/s) in the presence of squalene.

Figure 5 shows the average behavior of the curves of Figure 4 for $C_{12}$ and Br-$C_{12}$ along with the ones corresponding to mixtures of these oils with 15% squalene. These curves were obtained dividing the data in sets of 1 hour, and calculating the average value of the radius during those periods of time.

Since the radius is a sound measurement of the overall destabilization process, it is clear that the Br-$C_{12}$ emulsion is more unstable than the one of $C_{12}$. However, if enough squalene is added to each emulsion in order to decrease the rate of occurrence of Ostwald ripening significantly, the curves of Br-$C_{12}$ and $C_{12}$ coincide. This suggests that: a) Ostwald ripening is one of the main mechanisms of destabilization of these emulsions, and b) the effect of gravity is negligible in comparison to the influence of Ostwald ripening. Moreover, if the rate of change of the cube average radius ($V_{OR} = dR^3/dt$) of the emulsions is followed as a function of the amount of squalene in the oil phase, it is observed that an amount of 3% of $C_{30}$ is already sufficient to produce the same destabilization rates for both Br-$C_{12}$ and $C_{12}$ dispersions (Fig. 6). However, $V_{OR}$ decreases progressively with the percentage of $C_{30}$ until 10%, indicating that the effect of Ostwald ripening is significant until that threshold is reached. Above this percentage, further addition of squalene does not produce additional stabilization (Figure 6). These results agree with the fact that in the presence of a sufficient amount of insoluble oil in the dispersed phase, a balance between the Kelvin effect (which promotes Ost-wald ripening) and the Raoult effect (that strives against it) is achieved [Kabalnov, 2001]. Gravity produces only small differences between the pure systems, but for the mixed systems its influence is negligible. In other words, in the presence of Ostwald ripening gravity seems to reinforce the destabilization of the emulsions separating away drops of different size. When the ripening is suppressed the influence of gravity passes unnoticed at least during the measurement of the flocculation rate.

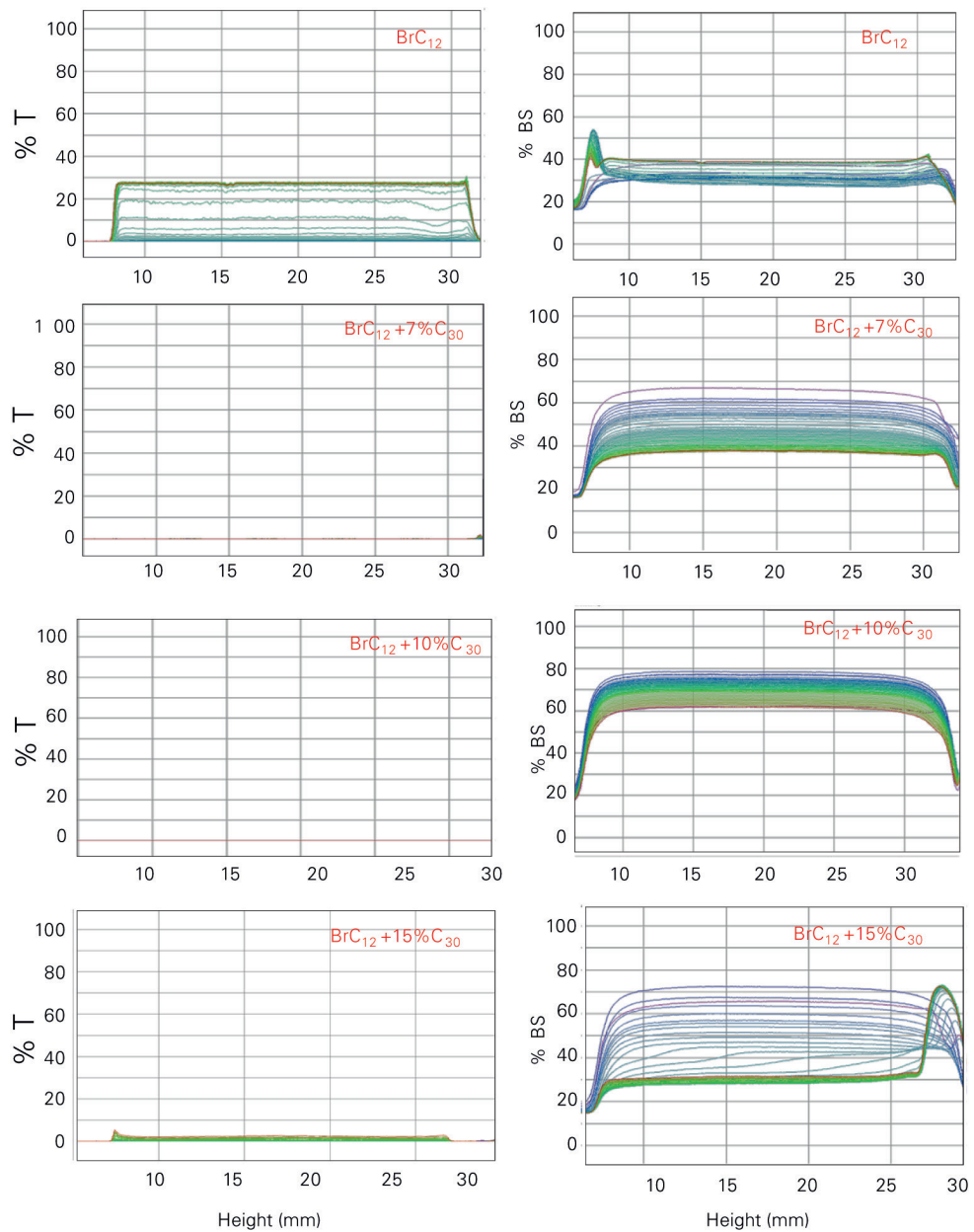

Figure 3: Temporal evolution of the percentage of Transmittance (%T) and Backscattering (%BS) of light scattered by drops of (bromo-dodecane (Br-$C_{12}$) + squalene ($C_{30}$)) suspended in water as a function of their height in the sample vessel.

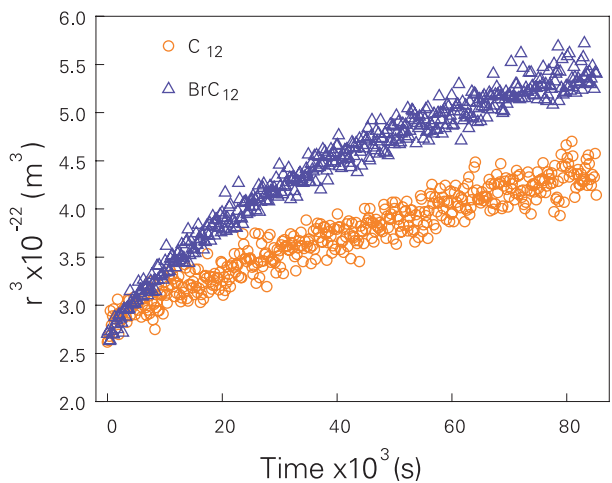

Figure 4: Variation of the cube average radius as a function of time for dodecane ($C_{12}$) and bromododecane (Br-$C_{12}$) in water nanoemulsions. Actual data as it comes directly from the Brookhaven Goniometer.

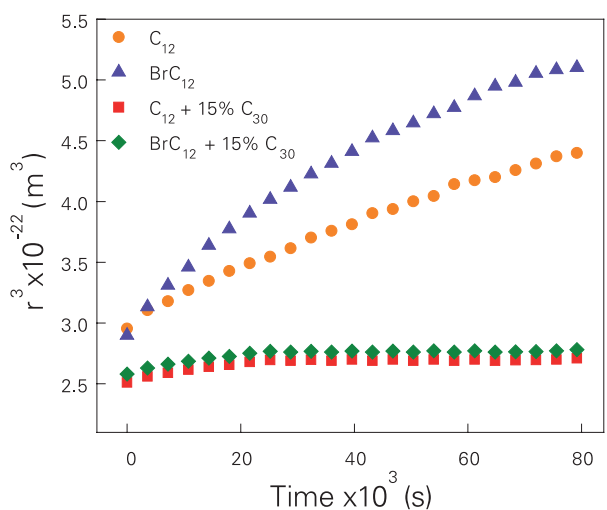

Figure 5: Variation of the cube average radius as a function of time for aqueous nanoemulsions of dodecane ($C_{12}$), bromododecane (Br-r$C_{12}$), ($C_{12}$ + 15% squalene), and (Br-$C_{12}$ + 15% squalene). The points are the result of an appropriate sampling of the data in order to illustrate the underlying behavior of the radius as a function of time time (see text).

Table 1: Average "initial" and "terminal" slopes of $R^3$ vs. t for the nanoemulsions of Fig. 5.

| SYSTEM | $dR^3/dt$ (m³/s) (0 < t < 3 h) | $dR^3/dt$ (m³/s) (3h < t < 24 h) |
|---|---|---|
| $C_{12}$ | $(4.2 \pm 0.2) \times 10^{-27}$ | $(1.5 \pm 0.4) \times 10^{-27}$ |
| Br-$C_{12}$ | $(5.8 \pm 0.5) \times 10^{-27}$ | $(1.8 \pm 0.1) \times 10^{-27}$ |
| $C_{12}$ + 15% $C_{30}$ | $(2.3 \pm 0.2) \times 10^{-27}$ | $(2.8 \pm 0.3) \times 10^{-29}$ |
| Br-$C_{12}$ + 15% $C_{30}$ | $(2.9 \pm 0.1) \times 10^{-27}$ | $(3.3 \pm 1.5) \times 10^{-29}$ |

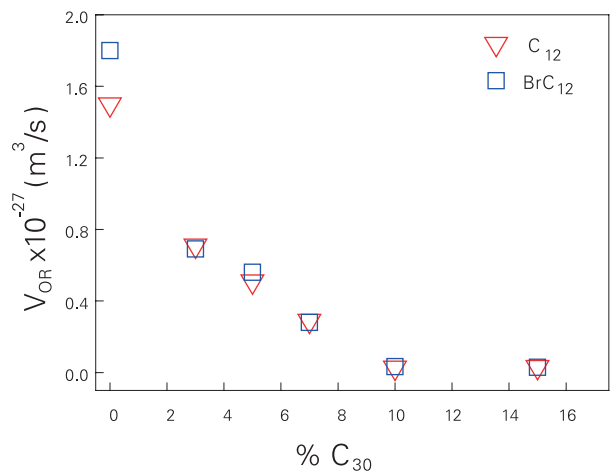

Figure 6: Ostwald ripening rates ($dR^3/dt$) for "finite-buoyancy" ($C_{12}$) and "null-buoyancy" (Br-$C_{12}$) nanoemulsions as a function of the squalene percentage.

Figure 7 illustrates the variation of the absorbance of the pure dodecane nanoemulsion as a function of time for a set of different temperatures. Since the amount of scattered light is a function of the size of the drops, an increase in the average size of the drops of the emulsion increases its adsorbance. The same thing occurs if the particles aggregate. As shown in Fig. 7, there is a progressive increase in absorbance as a function of time for all temperatures. The higher the temperature the greater the destabilization rate of the system.

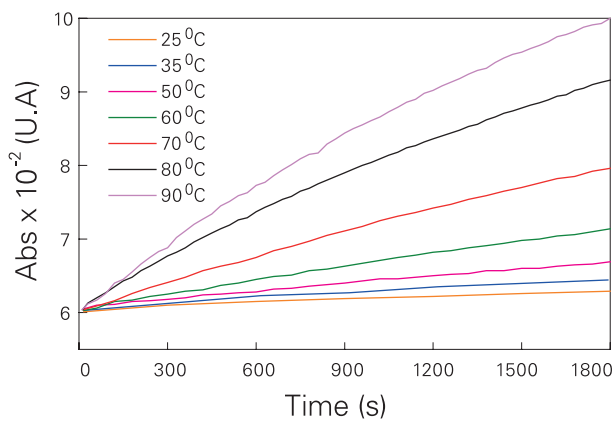

Figure 7: Evolution of the absorbance as a function of time corresponding to a set of dodecane-in-water nanoemulsions stabilized with Brij30 at different temperatures.

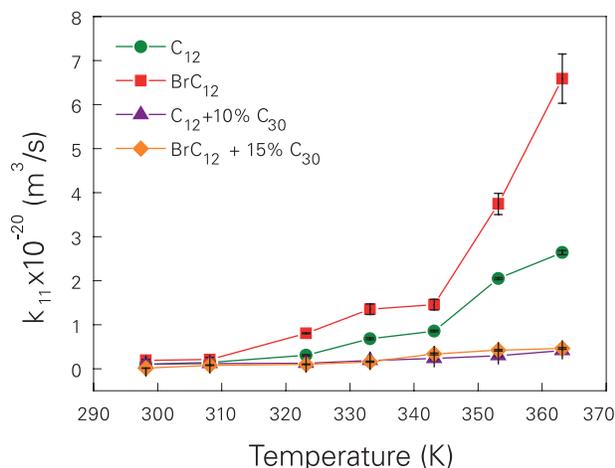

Figure 8: Comparison of the rates of doublet formation ($k_{11}$) for finite buoyancy, null buoyancy, and 10% and 15% mixtures of these components with squalene ($C_{30}$). Since the solubility of $C_{12}$ is lower than the one of Br-$C_{12}$ an amount of 10% of squalene is enough to delay the process of ripening considerably.

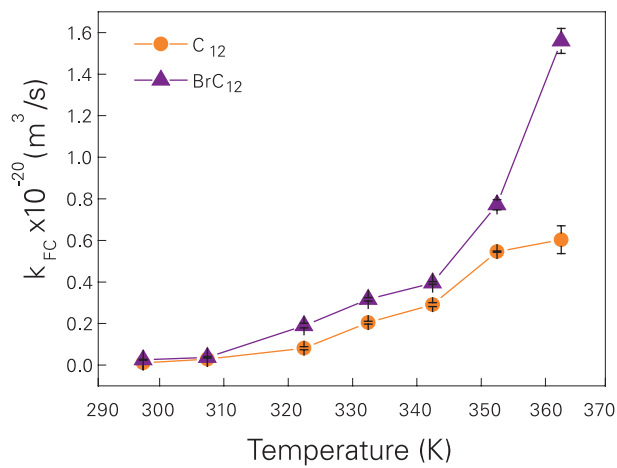

Figure 9: Constant-kernel flocculation-coalescence rates ($k_{FC}$) resulting from the adjustment of Eq. (7) to the experimental turbidity of oil-in-water emulsions as a function of time. (a) dodecane ($C_{12}$). (b) Br-$C_{12}$.

In regard to aggregation, the initial derivative of the adsorbance: $(dAbs/dt)_0$, can be used to calculate the rate of doublet formation following Eq. (6). Figure 8 shows that the flocculation rates $k_{11}$ increase with temperature and substantially decrease with the amount of squalene. In the absence of orthokinetic aggregation, flocculation occurs mainly by collisions between particles induced by Brownian motion (perikinetic aggregation). Brownian motion is directly proportional to the temperature, so an increase of $k_{11}$ with T is observed. In fact, the value of the flocculation rate, $k_f$, proposed by Smoluchowski for particles colliding exclusively as the result of Brownian motion is: $4k_BT/3\eta$, where $k_B$ is the Boltzmann constant. Additionally, it is well established that the increase of temperature promotes the desolvation of the ethylene oxide (EO) chains of non ionic surfactants. The EO-chains contract as a function of temperature, preventing their mixing with the chains of other drops during the process of aggregation. Thus, the repulsive osmotic contribution which characterizes the steric potential diminishes. Accordingly, the repulsive potential should decrease as the temperature increases, and as a consequence, the values of $k_{11}$ should increase with the temperature.

Figure 8 confirms our previous findings in the sense that the rate of doublet formation is different for finite buoyancy ($C_{12}$) and null buoyancy (Br-$C_{12}$) systems as long as Ostwald ripening is present. The same trend is observed in the case of the overall flocculation rate, $k_{FC}$ (Figure 9). Once the Ostwald ripening influence is reduced by adding squalene ($C_{12}$ + 10%$C_{30}$ and Br-$C_{12}$+15%$C_{30}$), the gravity does not make any significant difference in the flocculation rate. But in the presence of Oswald ripening, there is a sizeable difference between the flocculation rates of $C_{12}$ and Br-$C_{12}$. Notice, that the addition of squalene to the oil lowers the average density of the drops, but even in this case, the values of $k_{11}$ for $C_{12}$ + 10% $C_{30}$ and Br-$C_{12}$ + 15% $C_{30}$ almost coincide (Fig. 8).

## CONCLUSIONS

Although, the emulsions studied in this work should not cream until 5 months according to theoretical considerations, the formation of cream is observed after 24 hours. However, the increase in the average size of the drops is basically due to Ostwald ripening. When Ostwald ripening is suppressed, it is confirmed that the effect of gravity over the evaluation of the flocculation rate is negligible. When the mechanism of ripening is present, a sizeable difference between the flocculation rate of dodecane and bromo-dodecane in-water emulsions is observed. Hence, it is the ripening process which influences the evaluation of the flocculation rate, not the effect of buoyancy.